\documentstyle[11pt,newpasp,twoside,graphicx]{article}
\begin{document}
\title{Simulations of Sunyaev-Zel'dovich maps and their 
applications}
 \author{J. Delabrouille$^1$, J.-B. Melin$^1$ and J.G. Bartlett$^{1,2}$}
\affil{1. PCC - Coll\`ege de France, 11, place Marcelin Berthelot, 75231 
Paris cedex 05\\
2. Physics Department, Universit\'e Paris 7--Denis Diderot}

\begin{abstract}
We describe a fast method to simulate in a semi--analytical way 
consistent maps of the thermal, kinetic and polarised 
Sunyaev--Zel'\-do\-vich effect, featuring both cluster spatial 
correlations and large--scale velocity flows.
\end{abstract}

\section{Introduction}

There are a large number of applications for simulated maps of the 
Sunyaev-Zel'dovich (SZ) emission of galaxy clusters.  Such maps are 
being used, for instance, to assess the capabilities of instruments 
dedicated to the SZ observations, or for the development of data 
analysis tools.  They can also be used to constrain models of 
structure formation and/or cosmological parameter sets by comparison 
to real data.

There already exists a large number of tools to simulate SZ maps (Bond 
et al.  1996, Aghanim et al.  1997 , da Silva et al.  2000, Kay et al.  
2001, Kneissl et al.  2001, da Silva et al 2001).  Such tools range 
from simple uncorrelated Poisson distributions of idealised clusters 
to sophisticated hydro-codes simulating the complex physical process 
of structure formation, including non-linear gravity and some 
additional ingredients, such as magnetic fields, dissipation\ldots 
While the former may be over simplistic for a range of applications, 
the latter require large computing facilities and extensive computer 
time, and do not illuminate simply the influence of the code inputs on 
the final results.  In the approach presented here, we develop a code 
for simulating SZ maps that is fast enough to permit numerous 
Monte--Carlo simulations in a wide range of cosmological scenarios, 
while being accurate enough to correctly account for all three SZ 
effects (thermal, kinetic and polarised) and their distribution in 
both real and velocity space (e.g., bulk flows) according to the 
underlying spectrum of density fluctuations $P(k)$.

\section{The SZ effect and galaxy clusters}

Observations of galaxy clusters are useful as a probe of cosmology, 
and observations of the SZ effect in particular present various 
advantages in this light.  Most notably, as galaxy clusters are the 
largest bound objects in our Universe, their distribution as a 
function of redshift is extremely sensitive to the history of 
large--scale structure formation.  With its distance independence, the 
SZ effect is ideal for tracing this history out to large redshifts, 
and thereby for providing stringent constraints on models.  The SZ 
effect, although for long below attainable instrumental sensitivity, 
now represents an established and unbiased tool for the detection of 
distant clusters and for the estimation of their hot gas content.  In 
particular, the Planck mission, to be launched by ESA in 2007, will 
produce an all--sky catalog of several tens of thousands of clusters, 
which will be used to constrain scenarios of structure formation and 
cosmological parameters, for example, through cluster count tests.  In 
addition, SZ-selected cluster catalogs will be useful to constrain the 
two-point correlation function of clusters, and hence the matter power 
spectrum $P(k)$ on the largest scales.

The kinetic and polarized SZ effect, although much more difficult to 
detect than the thermal SZ, also appear as extremely promising tools, 
allowing the determination of both radial and transverse velocities 
for clusters.  Although detecting the polarised SZ effect still 
appears only marginally possible (Audit \& Simmons 1999, Sazonov \& 
Sunyaev 1999), the measurement of bulk radial velocities through the 
kinetic SZ effect seems within the reach of upcoming instruments 
(Aghanim et al.  1997).  Both will become of ever greater importance 
for cosmology as instruments improve.

\subsection{Why a new code for simulating SZ effect maps}

A key issue in interpreting SZ observations is the understanding of 
the selection function for clusters and of the importance of 
measurement errors.  The assessment both requires testing methods for 
the extraction of the relevant SZ observables on simulated 
observations that accurately model the cluster population -- gas 
physics and distribution in both real and velocity space -- as well as 
instrumental effects and relevant astrophysical foregrounds.  The 
simulations should be performed over a wide range of cosmological 
scenarios.  As an example, one expects that in the near future direct 
constraints of cosmological parameters will be obtained from SZ 
cluster counts and redshift distributions.  Likelihood contours may be 
obtained using analytical predictions of cluster counts (or cluster 
distributions in the observable parameter space), but the approach 
must be tested using Monte--Carlo simulations with numerous 
realisations of the observations and subsequent data analysis.  
Clearly, this can not be done using full--scale structure formation 
codes simulating gravitational collapse, as such codes require days or 
weeks of computation on supercomputers.

Our goal is to simulate maps accurate enough that they reliably 
feature the physical properties that convey cosmological information 
(e.g., cluster counts, cluster correlations in real space, bulk 
flows), in computation times permitting one to test methods 
statistically on thousands of simulations over a wide range of 
cosmologies.

\subsection{The formation of cluster sized dark-matter haloes}

The formation of large haloes takes place through the gravitational 
instability of dark matter density fluctuations.  
Original density fluctuations (at the time of 
matter radiation decoupling), still well within the linear regime, 
grow through gravitational instability into the non-linear regime 
(when the density contrast $\delta = \delta \rho / \rho$ becomes of order 
unity).  Collapsed haloes of dark matter appear where 
gravitational attraction overcomes  the 
expansion and the density contrast becomes larger than some limit.  The 
Press-Schechter (PS) formalism (1974) predicts the average number of 
such dark haloes as a function of mass, $M$, and redshift, $z$, for  
Gaussian fluctuations:
\begin{equation}
\frac{dN}{d\ln M dV_{c}} = \sqrt{\frac{2}{\pi}} \frac{\rho_{0}}{M} \nu \left | 
\frac{d\ln \sigma(0,M)}{d\ln M} \right | \exp(-\nu^2/2)
\end{equation}  
where $\nu = \delta_{c}(z) / \sigma(z,M)$,  $\rho_{0}$ is the present 
average matter density, $dV_{c}$ the covolume element, $\delta_{c}(z) 
\simeq 1.68$ the critical density contrast necessary for the collapse, 
and $\sigma(z,M)=D_g(z)\sigma(0,M)$ is the linear density fluctuation 
amplitude evolving according to the linear growth factor $D_g(z)$
($\equiv 1$ at $z=0$).
The fluctuation amplitude $\sigma(z,M)$ is related to the  power spectrum 
$P(k)$ by:
\begin{equation}
    \sigma^2(z,M) = D_g^2(z) \frac{1}{2\pi^2}\int_{0}^{\infty} k^2 P(k) W(k)^2 dk
\end{equation}  
\noindent where $W(k)$ is the Fourier transform of a 3D 
tophat smoothing window 
function of size $R = (3M/4\pi\rho)^{1/3}$.  The power spectrum 
$P(k)$, however, enters this formalism only through its integral value 
(the variance of the fluctuations);  therefore, the formalism cannot 
account for the spatial correlation of clusters.  Simulations based 
simply on PS result in maps that can only reproduce the 
predicted number of collapsed halos as a function 
of redshift, and not the cluster two--point correlation 
compatible with the underlying $P(k)$.

     The power spectrum uniquely defines the two--point correlation
properties of the density perturbation field, $\delta$.  Kaiser (1984)
was the first to suggest that clusters trace the density field in
a biased fashion, preferentially forming in overdense regions:
\begin{equation}
N_{\rm clusters} \propto 1+b\delta
\end{equation}
where the bias parameter $b(M,z)$ is in general a function
of both mass and redshift.  This behavior is explained by the
fact that galaxy clusters form from only the rarest, highest
density peaks in the density field, and it is borne out
by large N--body simulations; for example, Sheth et al. (2001) 
give an analytic expression for $b(M,z)$, based on earlier
work of Mo \& White (1996), that matches the clustering 
seen in simulations. 
We note in passing that, also from studies of clustering in large N--body 
simulations, the Press-Schechter formula is known to somewhat over-predict
the number density of small mass dark haloes and to
underestimate large haloes.  These studies have lead to 
improved expressions for halo number density  (e.g., Jenkins et al. 2001).

\subsection{Cluster velocities}
  
Cluster velocities are needed for the simulation of both kinetic SZ  
(radial part) and polarised SZ maps (transverse part). In 
the simplest simulation approach, the velocity of each individual cluster 
can be drawn at random from a 3D Gaussian probability distribution with
{\em rms} proportional to the square root of the integrated velocity 
power spectrum:
\begin{equation}
    \sigma_{v}(z,M) = a(z) H(z) f(\Omega_{m},\Lambda) 
    \left [ \frac{1}{8\pi^3} \int_{0}^{\infty} 4\pi D_g^2(z)P(k) W(k)^2 dk
    \right ]^{1/2}
\end{equation}  
with $a(z)=1/(1+z)$ and $H(z)=H_{0}\sqrt{\Omega_{m}(1+z)^3 + 
(1-\Omega_{m}-\Lambda)(1+z)^2 + \Lambda}$.  The function $f$ is 
defined as $f\equiv d\ln D_g(z)/d\ln a$.  Such an approach results in 
velocities which have the proper magnitude, but which are not 
correlated from one cluster to the next (i.e., no bulk flows).  For 
sensitivity reasons, and because of confusion due to the primordial 
CMB fluctuations (and other foregrounds), upcoming experiments will 
not be able to measure individual cluster velocities (Haehnelt and 
Tegmark 1996, Aghanim et al.  2001).  An estimate of the ability of 
instruments like Planck to measure peculiar velocities on large scales 
can be obtained by extrapolating the expected accuracy on a single 
cluster to a larger number of clusters moving together in a bulk flow 
(Aghanim et al.  2001).  However, a more accurate study of this 
approach and its uncertainties requires simulations that correctly 
model large--scale velocity correlations.  We detail such a method in 
the next section.

\begin{figure}
\begin{center}
\resizebox{6cm}{4.5cm}{\includegraphics{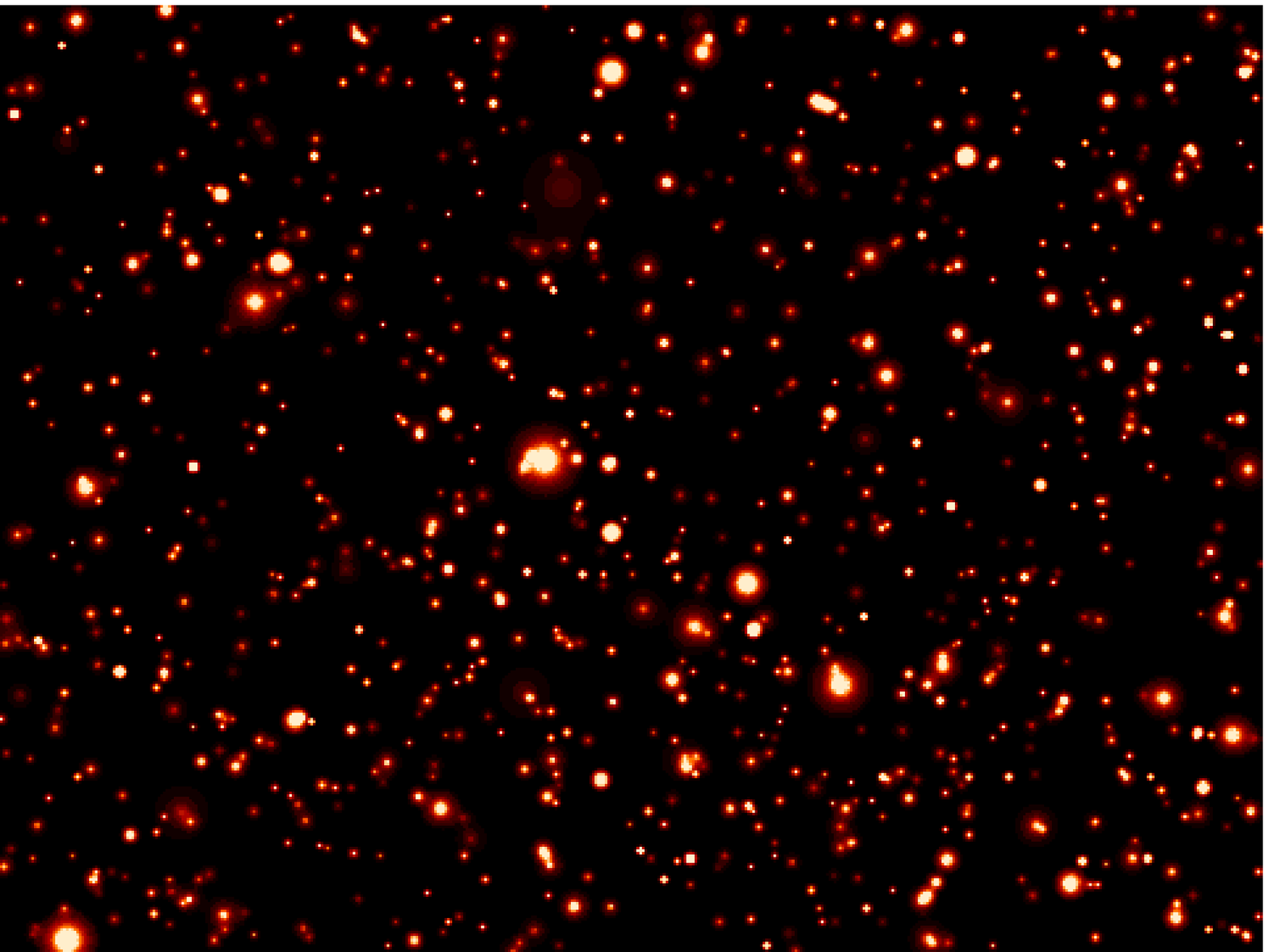}}\hspace*{1cm}
\resizebox{6cm}{4.5cm}{\includegraphics{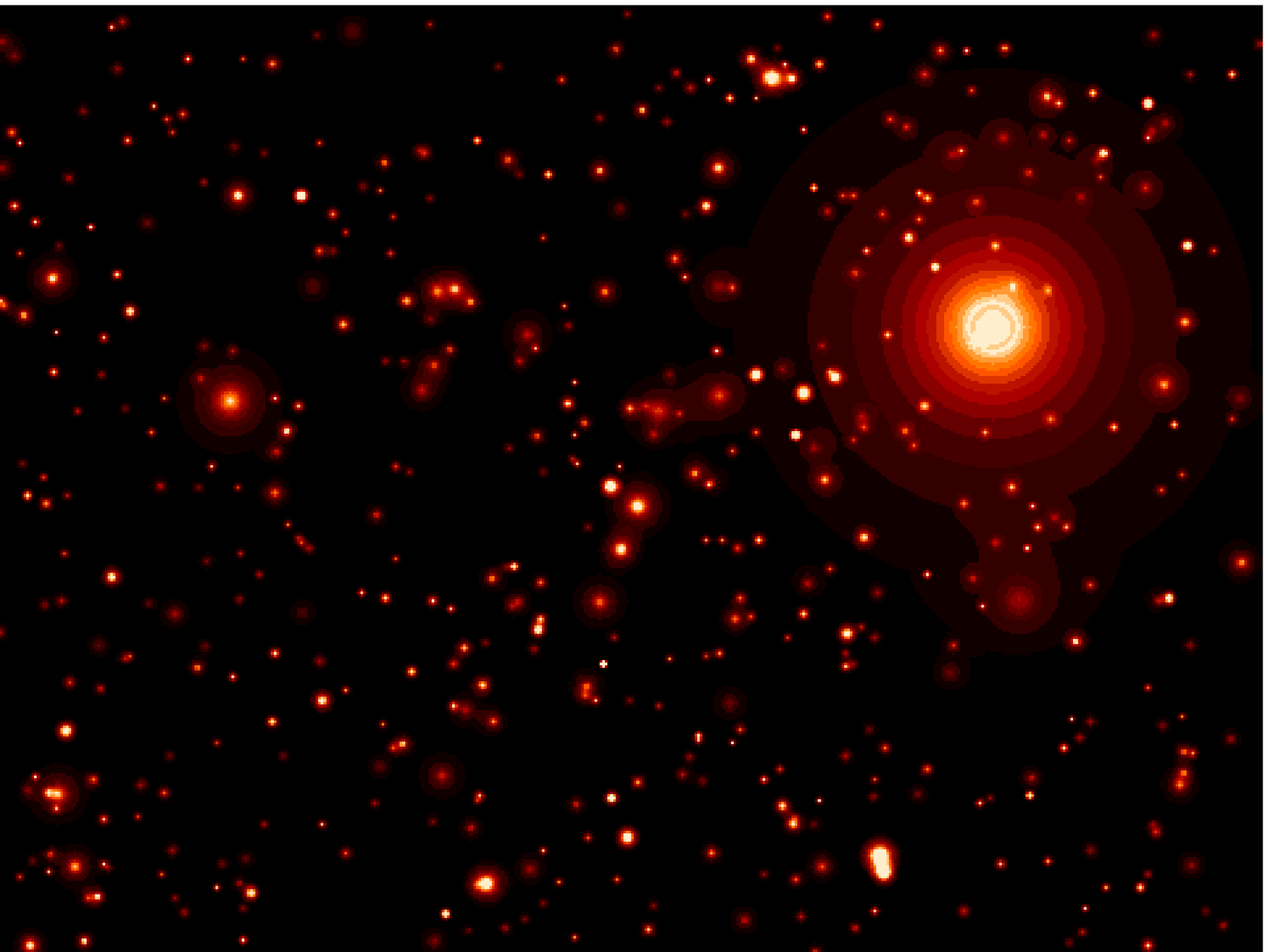}}
\end{center}
\caption{Thermal SZ map of 3 by 4 square degrees for a universe with 
$\Omega_{m}=0.3$, $\Lambda = 0.7$, $h=0.65$, $\Omega_{b}=0.05$ (left 
panel), and with $\Omega_{m}=1$, $\Lambda = 0$, $h=0.5$, 
$\Omega_{b}=0.07$ (right panel).  Note that because of the different 
baryon fraction, the $\Lambda$ universe clusters are brighter in SZ 
(respective color scales have been chosen such that they range from 
$y=0$ to $y=4\times 10^{-5}$ on the left panel , and from $y=0$ to 
$y=1.5\times 10^{-5}$ on the right one, with higher values saturated.}
\end{figure}

\section{Simulation method}

\subsection{The cluster spatial distribution}

Since the power spectrum of density fluctuations, $P(k)$, uniquely 
defines the density field responsible for the formation of halos and 
their clustering, we start with a random realisation of the density 
contrast in a 3D comoving box, with the user placed at one end.  The 
size of the box is matched to the size of the required map (angular 
extent).  For a 3 degree--by--3 degree map, typical comoving box sizes 
(extending to a redshift of 5) are $600 \times 600 \times 6000$ Mpc, 
divided in resolution cells of about 20 Mpc.  To account for the 
different redshifts seen by the observer while looking through the 
box, the density field is scaled by the linear growth factor, 
$D_{g}(z)$, over slices of redshift (distance from the user).  A 
random catalog within the box is first constructed by drawing from a 
Poisson distribution with average given by the PS law (or 
Sheth-Tormen, or Jenkins et al., as appropriate, up to the user of the 
simulation tool) the number of clusters in a grid of mass bins for 
each redshift shell.  The average can be slightly modified to take 
into account the fact that the particular redshift shell might be 
slightly overdense or underdense with respect to the ensemble average.  
For each shell, we then distribute the clusters in $(\theta,\phi)$ 
with a probability proportional to $1+b\delta$, where $\delta$ is the 
density field scaled to the redshift of the shell and $b$ the 
clustering bias.  In this way, we construct a cluster distribution 
compatible with the underlying $P(k)$ (within the resolution of the 
code).

\subsection{Cluster peculiar velocities}

The peculiar velocity of each cluster is found using 
linear theory applied to the density field realisation, 
$\delta \rho / \rho$. With the assumption that the peculiar 
velocity field is irrotational, as expected in most scenarios of 
structure formation, the velocity field can then be obtained from the 
Newtonian gravitational potential through
\begin{equation}
    v = -\frac{2}{3} \frac{f(z)}{\Omega_m(z)H(z)} \frac{\nabla \Phi}{a}
\end{equation}  
where the gravitational potential $\Phi$ is connected to $\delta = 
\delta \rho / \rho$ by
\begin{equation}
    \nabla^2 \Phi = 4 \pi G a^2 \delta\rho
\end{equation}  

\begin{figure}
\resizebox{13cm}{9.75cm}{\includegraphics{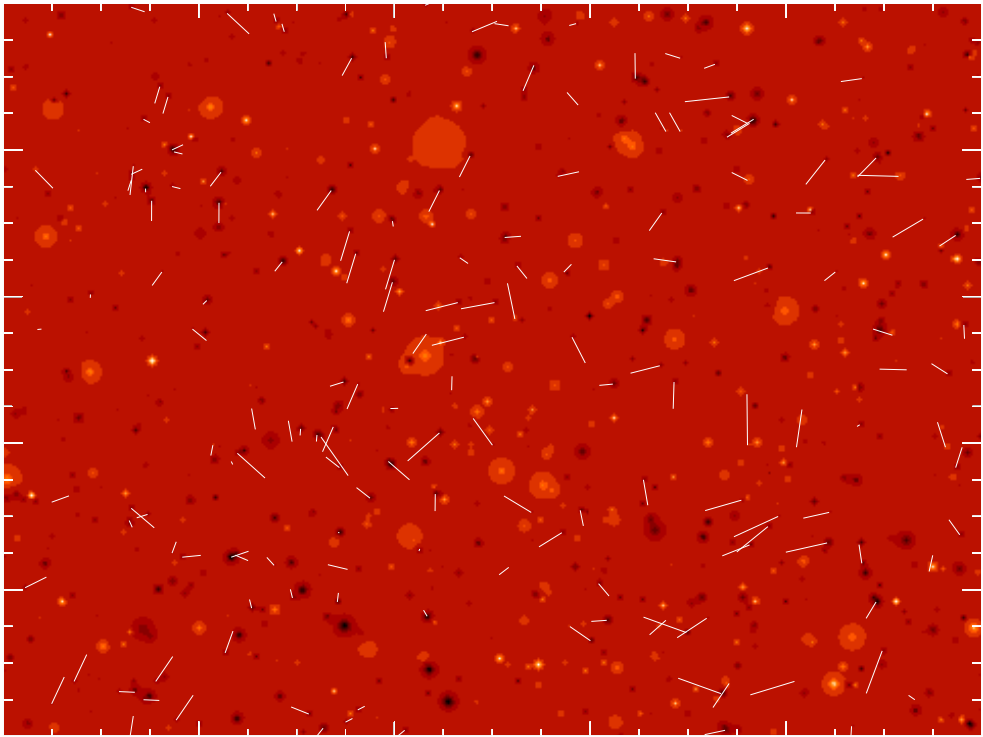}}\hspace*{1cm}
\caption{Velocity map of 3 by 4 square degrees.}
\end{figure}
\subsection{Cluster gas properties}

The physical modeling of individual clusters is a complicated issue, 
in principle calling into play a variety of physical processes.  As 
spectacularly illustrated by the new X--ray satellites XMM--Newton and 
Chandra, clusters are not simple objects, but rather display complex 
substructures and inhomogeneities.  For large maps which will be 
observed at low angular resolution (as will be the case for the Planck 
mission, for instance), these substructures are lost and thus not very 
important for the final properties of the map.  For resolved clusters, 
on the other hand, proper physical modeling of cluster substructure is 
an important issue.

In our current simulations, we adopt a simple, self--similar cluster 
model based on an isothermal $\beta$ model:
\begin{equation}
    n_{e}(r) = n_{e}(0)\left (1+ \left (  \frac{r}{r_{c}}\right )^2 
    \right)^{-3\beta/2}
\end{equation}
for the density profile, and a gas temperature--mass
relation normalized to numerical simulations (Evrard et al. 1996):
\begin{equation}
    T_{e} = 6.8 h^{2/3} \left ( \frac{M}{10^{15} M_{\odot}} 
    \right )^{2/3}\left(\frac{\Omega_m\Delta_{NL}(z)}{178}\right)^{1/3} 
    (1+z) \; {\rm keV}
\end{equation}
where $\Delta_{NL}(z)$ is the non--linear density contrast on collapse
(a weak function of both $\Omega$ and $\Lambda$ that equals 178 in
a critical universe).  The virial radius is given by (e.g., 
Bartlett 1997)
\begin{equation}
    r_{v} = 1.69 h^{-2/3} \left ( \frac{M}{10^{15} M_{\odot}} \right )^{1/3}
    (1+z)^{-1} \left ( \frac{178}{\Omega_{m}\Delta_{NL}(z)} 
    \right ) ^{1/3} \; {\rm Mpc}
\end{equation}
Another possible approach is to replace this simple cluster model 
by maps of the Comptonization parameter $y$ and/or the optical 
depth $\tau$ obtained in detailed simulations of individual clusters, 
as done by Kneissl et al. (2001).

\section{Examples of maps obtained with the software}

In Figure 1 we show $ 3\times 
4$ degree$^2$ SZ maps simulated with our code for two different 
cosmologies, one with 
($\Omega_{m}=0.3$, $\Lambda = 0.7$, $h=0.65$, $\Omega_{b}=0.05$), left
panel, and 
one with ($\Omega_{m}=1$, $\Lambda = 0$, $h=0.5$, $\Omega_{b}=0.07$),
right panel.
The visual difference between the two maps is striking, 
with more high redshift clusters in the $\Lambda$--model 
than in the critical model (Barbosa et al. 1996). 

As another example, we show in Figure 2 a peculiar velocity map where 
radial velocities are coded in color (the map is one of radial 
velocity times optical depth, $\beta_{r} \tau$).  Transverse 
velocities for clusters between redshifts 0.8 and 1 are indicated by 
the small arrows starting from the center of each cluster.  The 
tangential velocities in this limited redshift slice demonstrate clear 
correlations on large scales.  Maps of the polarised emission are 
straightforwardly obtained from the optical depths and transverse 
velocities, but are not shown here due to a lack of space.

\section{Characteristics of the code}

The software can produce simulations for critical, open or closed 
CDM-like models, with or without a cosmological constant. A future  
development will be the inclusion of an arbitrary equation--of--state 
for the dark energy term (Quintessence).  Presently, our software can 
only produce small tangential maps of limited fractions of the sky,
although we are currently working on full-sky extensions.
The software is implemented in the IDL programming language, and 
requires only a couple minutes on a modest PC to produce a map
of several degrees squared.  


\section{Conclusion}

We have written a fast code for the simulation of SZ maps (thermal, 
kinetic and polarised) based on a simple model of cluster physics and 
on Monte Carlo realizations of the mass function and the linear 
density field.  The code produces a cluster distribution and 
large--scale peculiar velocities consistent with the underlying power 
spectrum, $P(k)$.  The software can produce maps for any kind of open, 
closed or critical universe, with or without a cosmological constant.  
With the expectation of data from the new generation of SZ dedicated 
instruments, this will be a useful tool for a wide range of 
applications, from optimizing observations strategies and 
devising/testing data processing methods, to constraining cosmological 
scenarios with real data.

\end{document}